\documentclass[lettersize, draftcls, onecolumn]{IEEEtran}
\usepackage{cite}
\usepackage{graphicx}
\usepackage{amsmath,amssymb,amsfonts}
\usepackage{algorithmic}
\usepackage{array}
\usepackage[caption=false,font=normalsize,labelfont=sf,textfont=sf]{subfig}
\usepackage{textcomp}
\usepackage{stfloats}
\usepackage{xcolor}
\usepackage{url}
\usepackage{lineno}
\usepackage{hyperref}
\hypersetup{
    colorlinks=true,
    linkcolor=blue,
    filecolor=magenta,      
    urlcolor=cyan,
}

\renewcommand{\figureautorefname}{Figure}
\newcommand{\figuresautorefname}{Figures}
\renewcommand{\tableautorefname}{Table}

\renewcommand{\sectionautorefname}{Section}

\def\BibTeX{{\rm B\kern-.05em{\sc i\kern-.025em b}\kern-.08em
    T\kern-.1667em\lower.7ex\hbox{E}\kern-.125emX}}
\usepackage{balance}

\begin{document}

\title{Multi-Agent Digital Twinning for Collaborative Logistics: Framework and Implementation
\thanks{This work has been accepted to the Journal of Industrial Information Integration for publication.}
}

\author{Liming~Xu, Stephen~Mak, Stefan~Schoepf, Michael~Ostroumov, and Alexandra~Brintrup
\thanks{L. Xu (Email: lx249@cam.ac.uk), S. Mak, S. Schoepf, A. Brintrup are with Supply Chain AI Lab, Institute for Manufacturing, Department of Engineering, University of Cambridge, Cambridge, UK.}
\thanks{M. Ostroumov is with Value Chain Lab, London, UK}
}

\markboth{Journal of \LaTeX\ Class Files,~Vol.~18, No.~9, September~2020}%
{How to Use the IEEEtran \LaTeX \ Templates}

\maketitle
\IEEEpeerreviewmaketitle

%% Abstract
\begin{abstract}
Collaborative logistics has been widely recognised as an effective avenue to reduce carbon emissions by enhanced truck  utilisation and reduced travel distance. 
However, stakeholders' participation in collaborations is hindered by information-sharing barriers and absence of integrated systems.
We, thus, in this paper addresses these barriers by investigating an integrated platform that facilitates collaboration through the integration of agents with digital twins. 
Specifically, we employ a multi-agent system approach to integrate stakeholders and physical assets in collaborative logistics, representing them as agents.
We introduce a loosely-coupled system architecture that facilitates the connection between physical and digital systems, enabling the integration of agents with digital twins. 
Using this architecture, we implement a prototypical testbed.
The resulting testbed, comprising a physical environment and a digital replica, is a digital twin that integrates distributed entities involved in collaborative logistics. 
Its effectiveness on integrating both physical and digital, stationary and mobile objects is demonstrated through a carrier collaboration scenario. 
This paper is among the few earliest efforts to examine the integration of agents and digital twin concepts in logistics sector and goes beyond the conceptual discussion of existing studies to the technical implementation of such integration. 
\end{abstract}


% Introduction 
\section{Introduction}\label{sec:introduction}
Transportation is the largest contributor to greenhouse gas emissions \cite{transport2022energy}.
Among various transportation modes, trucks are the second-largest emitter following cars and taxis.
However, they are currently utilised inefficiently, operating at only around 60\% of their weight capacity, with approximately 30\% of their travelled distance carrying no freight \cite{dft2020roads}.
Collaborative logistics has been widely recognised as an effective pathway to enhance truck utilisation \cite{cruijssen2007horizontal}\cite{mak2021coalitional}\cite{karam2021horizontal}. 
This approach involves carriers collaborating via coalition to collectively fulfil delivery requests, thereby achieving reduced total cost and travel distance through economies of scale.

Collaboration in private transportation, facilitated by centralised platforms such as UberPool\footnote{\url{https://www.uber.com/au/en/ride/uberpool/}}, has seen widespread adoption.
However, extending this model to the logistics sector remains relatively unexplored.
Two primary barriers, among others \cite{karam2021horizontal}, contribute to this challenge:
1) {\it Lack of Trusted Platforms}: Concerns about business secrecy {often discourage} carriers from sharing data with centralised platforms, despite the evident environmental and economic benefits.
2) {\it Lack of Information}: Carriers operating in close proximity often lack visibility regarding into whether they are transporting freight to similar destinations at overlapping times.
These barriers impede stakeholders' willingness to engage in collaborative efforts.
One promising approach to overcome these barriers is to establishing a trusted integrated environment where carriers can confidently share information and actively participate in collaboration.

%% Figure: carrier collaboration
\begin{figure}
    \centerline{\includegraphics[width=0.55\textwidth]{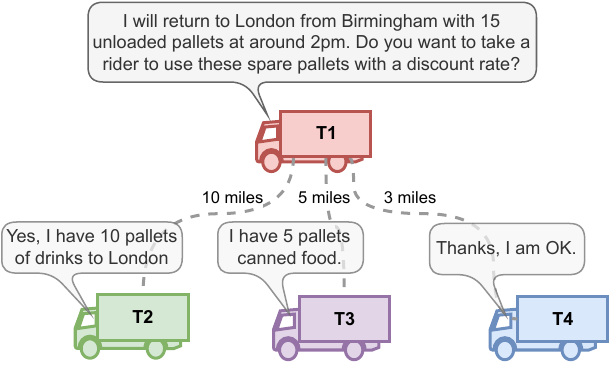}}
    \caption{Illustration of a horizontal collaboration scenario.}
    \label{fig:collaboration_illustration}
\end{figure}

However, as highlighted in \cite{karam2021horizontal}, studies investigating the implementation of such platforms are {\it scarce}, despite many studies \cite{adenso2014analysis}\cite{liu2010task}\cite{mak2023fair} proposed to address computational challenges, such as vehicle routing problems, in collaborative logistics.
These platforms could motivate carrier to participate in collaboration through better informed and tangible decision-making simulations on different operational scenarios, as reported in \cite{lindawati2014collaboration}.
Therefore, our paper aims to bridge this gap.
We design and develop an integrated environment for orchestrating collaborative logistics, leveraging the multi-agent system (MAS) approach and the digital twin technology.

Centralised platforms can help alleviate the barrier of lacking information, however, they may not ensure neutrality and trustworthiness. 
An ideal solution is a decentralised, distributed platform operated collectively by participants, where planning and decision-making are distributed rather than delegated to central authorities.
The MAS approach, often adopted for solving complex distributed problems, is well-suited for developing such platforms \cite{wooldridge1995intelligent}.

However, such platforms might not mitigate obstacles arsing from human factors, such as a lack of awareness regarding collaboration benefits or apprehensions about altering their business model \cite{karam2021horizontal}. 
To enable collaboration, carriers need to communicate with others, often those located nearby.
For better illustration, we present an illustrative carrier collaboration in \figureautorefname~\ref{fig:collaboration_illustration}.
As shown in \figureautorefname~\ref{fig:collaboration_illustration}, a truck with 15 unloaded pallets sends a message to nearby carriers to propose co-loading collaboration. 
If successful, this collaboration could reduce from five trips ({\tt T1}: one way, {\tt T2} and {\tt T3}: round trips) to just one one-way trip by {\tt T1}. 
This illustration showcase that large benefits can be achieved through carrier collaboration, despite its simplicity.
In this paper, we explore to create an integrated platform to facilitate and demonstrate such collaborations in logistics.

Recently, the digital twin technology has been widely discussed and explored in various domains for their potential in simulation, integration, monitoring, and maintenance \cite{grieves2018virtually, sharma2022digital}.
While the MAS approach can model the collaborative logistics problem
, it cannot provide visibility into and transparency of the operations of how collaborations are achieved, which is critical to create a trusted platform.
Therefore, we incorporate the digital twin technology into the development of this platform, in conjunction with the MAS approach, demonstrating instances of collaboration and mirroring the operation of the physical system in its virtual replica. 
This digital twin-based platform would effectively address both information sharing barriers and human-factor obstacles, enhancing stakeholders' trust and motivating their collaboration.
Although the conceptual integration of agents with digital twins has been explored in areas such as smart cities \cite{clemen2021multi} and healthcare \cite{croatti2020integration}, this integration in collaborative logistics is scarce and its corresponding technical implementation is absent.

This paper technically explores this integration in the domain of collaborative logistics, which may represent the first implementation of such integration in this domain.
We implement a testbed that consists of a physical environment and a digital system, providing a platform for testing collaborative scenarios. 
Moreover, this work may align with the recent development of the Physical Internet \cite{montreuil2011toward} initiative, which aims at enhancing transportation efficiency by building an ``Internet'' for transporting physical goods through the integration of physical and digital resources.

The main contributions of this paper can be summarised as follows:
\begin{itemize}
    \item We explore an agent framework for collaborative logistics, detailing agent decomposition,  organisation, and interaction. 
    
    \item We present a system architecture for integrating physical and digital objects, which may also be suitable for creating other hybrid systems. 
    
    \item We implement a testbed, discuss the design of its both physical and digital components, and demonstrate its effectiveness in facilitating studies on collaborative logistics.
\end{itemize}

The rest of this paper is structured as follows. 
\sectionautorefname~\ref{sec:related_work} reviews related works. 
\sectionautorefname~\ref{sec:macl_framework} presents the agent framework. 
\sectionautorefname~\ref{sec:testbed} details the design and development of the integrated platform. 
\sectionautorefname~\ref{sec:discussion} discusses limitations and implications of this work.
Finally, \sectionautorefname~\ref{sec:conclusion} concludes this paper and briefly describes future work.

%% Related work
\section{Related Work}\label{sec:related_work}
This sections reviews related work, including collaborative logistics, multi-agent systems, and digital twins.

%%  Related work: collaborative vehicle routing
\subsection{Collaborative Logistics}
Collaborative logistics, sometimes known as horizontal collaborations in logistics, has {attracted} increasing attention in the past decade \cite{cruijssen2007horizontal}\cite{cruijssen2007joint}.
Among the various challenges within collaborative logistics, one of central problems is vehicle routing problems \cite{dantzig1959truck}\cite{gendreau2008metaheuristics}, specifically known as CVRPs.  
In CVRPs, carriers collaborate by sharing their delivery requests to optimise routes, thereby reducing total travel distance, collective costs, and achieving environmental and socio-economic goals \cite{gansterer2018collaborative}\cite{karam2021horizontal}.

Both centralised and decentralised approaches have been proposed to CVRPs in literature \cite{gansterer2018collaborative}. 
In centralised approaches, information may require collection from all participants by a central authority. 
Subsequently, optimisation methods such as metaheuristics \cite{adenso2014analysis} and greedy heuristics \cite{ergun2007reducing} can be employed by this central authority to identify optimal collaborations.
Centralised collaborations can achieve significant improvements, up to 30\% in synergy values \cite{cruijssen2007joint} \cite{montoya2016impact}, and \cite{soysal2018modeling}; however, their extensive information gathering raises concerns about business privacy and would undermines companies' commitment and trust on these centralised solutions.
Therefore, many decentralised approaches \cite{berger2010solutions}\cite{cuervo2016determining}\cite{li2016adaptive}\cite{mak2023fair} have  been proposed to overcome these shortcomings.
While existing work mainly focus on addressing computation or optimisation challenges, studies on developing {\it integrated} platforms facilitating collaborative logistics remain limited \cite{karam2021horizontal}.

%% Related work: MAS approach
\subsection{Multi-Agent System Approach}
The MAS approach \cite{wooldridge1995intelligent} has been successfully applied in various domains, including transportation and logistics \cite{davidsson2005analysis}\cite{xu2021will}. 
This approach models distributed entities involved in collaboration as agents, achieving coherence through {interaction between these agents}.
Collaborative logistics, which involves distributed entities such as carriers, trucks, and shippers, is well-suited for adopting this approach to create the underlying architecture and facilitate connection among distributed parties.
Despite its widespread applications in transportation and logistics \cite{davidsson2005analysis}, its use in collaborative logistics is limited, especially in developing integrated systems combined with the digital twin technology.  
One of representative work is Dai and Chen \cite{dai2011multi}, where they presented a multi-agent framework for decentralised carrier collaboration, adopting an auction mechanism for managing transportation request outsourcing and acquisition. 
The MAS approach has also been applied to related fields, such as enabling autonomous supply chains \cite{xu2024implementing,xu2024towards}---an promising domain that has emerged in the era of large language models.

%% Related work: Digital Twin
\subsection{Digital Twin and Its Integration with Agents}
Digital twins are virtual representations of physical objects and processes, providing visibility into the operation of {their physical counterparts} and facilitating decision making through simulations of various scenarios \cite{tao2018digital}\cite{sharma2022digital}.
First publicly introduced in 2002 by Michael Grieves \cite{grieves2018virtually}, The digital twin technology has found applications in various domains, including aerospace \cite{glaessgen2012digital}, construction \cite{jiang2021digital}, manufacturing \cite{kritzinger2018digital}, supply chain \cite{sharma2022digital}, and logistics \cite{greif2020peeking}. 
However, their use in collaborative logistics, particularly their integration with agents is rare or non-existent.
Although conceptual discussions of this integration exist in smart cities \cite{clemen2021multi}, healthcare \cite{croatti2020integration}, its practical implementations are still lacking.
We thus in this paper further investigate this integration, developing a platform that integrates physical objects, agents, and digital twins for facilitating collaboration in logistics.
Through the integration of digital twin technology into multi-agent collaborative logistics, the digital system would enhance platform visibility, foster trust, and motivate participation in collaboration \cite{karam2021horizontal}.

%% Agent framework 
\section{Multi-Agent Collaborative Logistics Framework}\label{sec:macl_framework}
%% Figure: agent orgnisation
\begin{figure}
    \centerline{\includegraphics[width=0.65\textwidth]{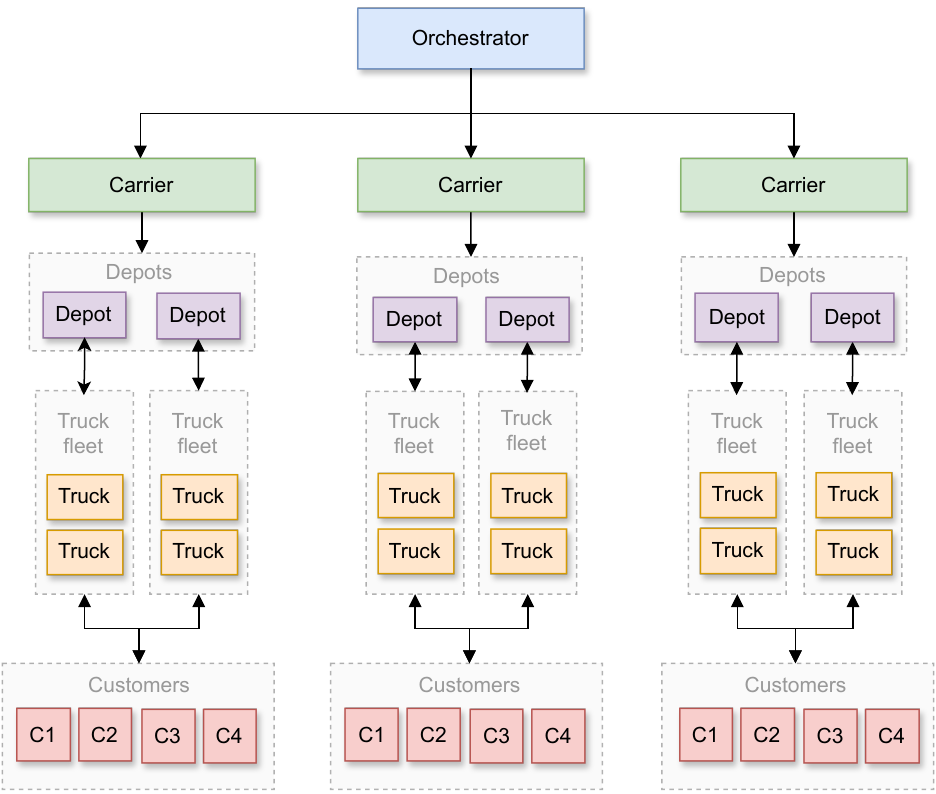}}
    \caption{An agent framework for carrier collaboration.}
    \label{fig:agent_organisation}
\end{figure}
This section presents the multi-agent collaborative logistics (MACL) framework, an agent framework that represents stakeholders in collaborative logistics as software agents that communicate through messaging.
This section aims to provide an overarching framework for enabling the creation of an integrated environment that facilitates testing of collaborative logistics scenarios.

%% Scenario
\subsection{Scenario}\label{sec:scenario}
Many transportation collaboration scenarios have been studied in existing literature  \cite{gansterer2018collaborative}.
This paper focuses on a specific scenario: carrier collaboration. 
In this scenario, carriers, which are the owners and operators of transportation equipment, work together to achieve reduced cost and enhanced efficiency.
Shippers, the shipment owners, are excluded from this scenario due to its carrier-oriented focus.
Therefore, this scenario involves a group of carriers, each owns a fleet of trucks, collaborating to fulfil their customers.
This scenario is described without considering technical details, allowing for both centralised and decentralised approaches to be employed for its realisation. 
It is worthy to note that real-world carrier collaboration scenarios can be more complex, with processes and corresponding stakeholders varying from case to case. 
This scenario is therefore a simplification, focusing on the most essential elements that may apply to most collaboration scenarios.

%% Agents
\subsection{Agent Decomposition}\label{sec:agents}
Based on {this scenario}, the agent framework consists of five primary agent types, detailed as follows:
\begin{enumerate}
    \item[1)] Orchestrator: 
    The orchestrator is a computational or algorithmic agent responsible for coordinating collaborations. 
    It conducts searches for collaborative solutions and optimises routes for delivery requests collected from carriers in the coalition. 
    
    \item[2)] Carrier: 
    Carrier agents represent the carriers who initiate delivery requests (shipments). 
    Carriers play an important role in collaboration in the given scenario, managing their transportation processes.
    They typically own multiple depots and operate a fleet of trucks. 
    Carriers aim to efficiently utilise their trucks to achieve both economical and ecological goals.
    
    \item[3)] Truck: 
    Truck agents represent the trucks owned by carriers.
    They receive transport requests and travel assigned routes to fulfil these requests. 
    
    \item[4)] Depot: 
    Depot agents represent the depots, acting as truck fleet bases. 
    Carriers may have one or more depots. 
    Trucks depart from their respective depots to fulfil requests and return to them after finishing assigned tasks.
    
    \item[5)] Customer: 
    Customer agents represent the recipients of the shipments --- customers.
\end{enumerate}

These agents are the main stakeholders in the MACL framework, each undertaking a specific role and collaborating with others to facilitate carrier collaboration.
These agents are BDI (Belief-Desire-Intention) agents \cite{rao1995bdi}, which are based on the BDI architecture developed by Michael Bratman \cite{bratman1987intention}.
While these section describes the roles of these agents, the next section delves into their organisation and corresponding behaviours during interaction.

%% Organisation and interaction 
\subsection{Agent Organisation and Interaction}\label{sec:interaction}
%% Figure: MACL interaction 
\begin{figure}[!t]
    \centerline{\includegraphics[width=0.75\textwidth]{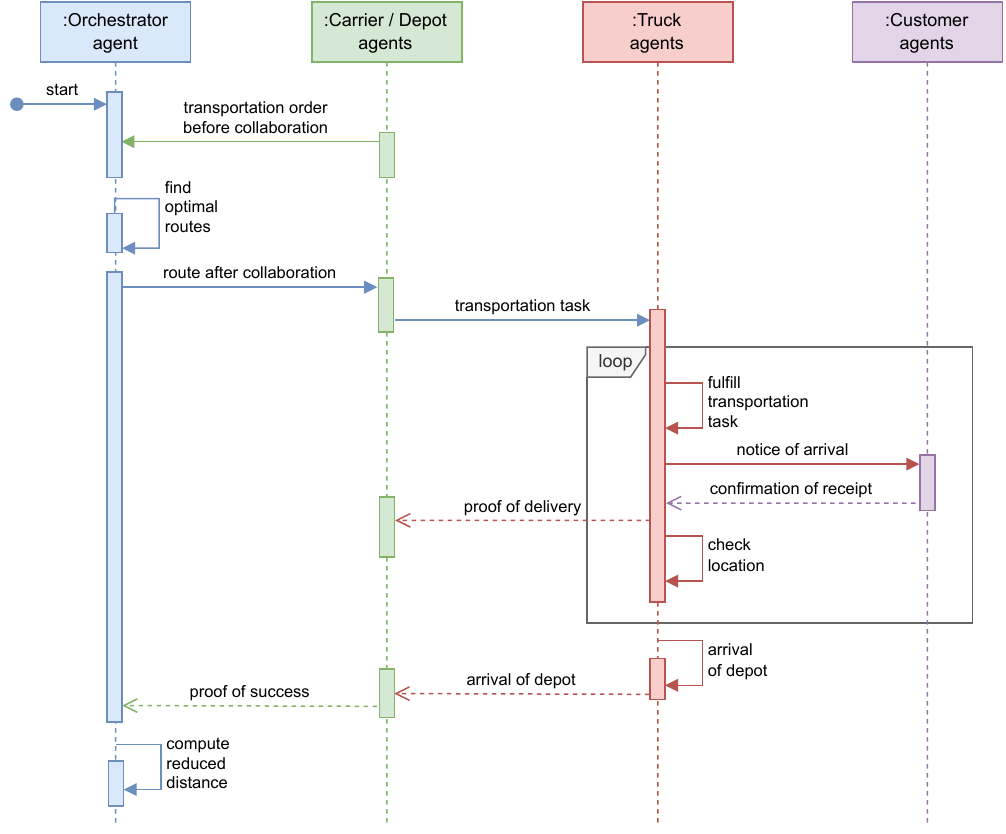}}
    \caption{Agent interaction sequence of the MACL.}
    \label{fig:agent_interaction}
\end{figure}
Agents interact to achieve goals and their interaction can be organised in various structures.
We present an agent organisation for the given scenario, as shown in \figureautorefname~\ref{fig:agent_organisation}.
For simplicity and clarity, the figure only includes a partial list of agents in each category.
{However, a real-world collaborative logistics scenario may include much more number of carriers and customers compared to the depiction} in \figureautorefname~\ref{fig:agent_organisation}.

As shown in \figureautorefname~\ref{fig:agent_organisation}, the proposed agent organisation adopts a hierarchical structure, where agents are conceptually organised as a tree-like structure. 
The orchestrator, a computational agent, coordinates collaborations among carriers in the coalition. 
Carriers select information to share with the orchestrator based on their own objectives. 
Carriers manage depots, each manages a fleet of assigned trucks.
Trucks interact with their respective depots when they depart from and return to them at the beginning and end of delivery tasks. 
Carriers handle transportation requests from shippers and then transport shipments to their recipients (i.e., customers). 
Direct communication between carriers and customers is disabled; instead, trucks engage with customers to manage shipment reception and confirmation. 
Customers, {which are loosely connected} to the system, stay reactive to incoming shipment arrival notifications.

As BDI agents, we further define their behaviours and interaction to enable collaboration in the given scenario.
These agents have either proactive or reactive behaviours, organised appropriately during their interaction. 
We present their interaction in \figureautorefname~\ref{fig:agent_interaction}, which illustrates a complete agent interaction process involved in a carrier collaboration. 
As shown in \figureautorefname~\ref{fig:agent_interaction}, agents interact with others through messaging, such as transportation orders, notice of arrival, and proof of delivery. 
Agents trigger corresponding behaviours to handle received messages with predefined handlers. Although, as per convention, these handlers are not displayed in this sequence diagram, corresponding handlers are defined for each agent to process all their incoming message. 
Specifically, the orchestrator agent starts to find optimal routes for the collection of transportation orders received from all collaborative carriers. 
When carrier agents receive optimal delivery routes from the orchestrator agent, they send transportation tasks to their truck agents for fulfilment.
After receiving transportation tasks with given routes, truck agents start their journey for delivery fulfilment. 
As depicted in the embedded frame of \figureautorefname~\ref{fig:agent_interaction}, the fulfilment journey involves a loop process: trucks navigate, continuously monitoring their real-time locations, and send notice-of-arrival message to their customers upon arrival.
This process continues until the truck fulfils all delivery orders and arrives at its depot. 
Finally, the orchestrator agent calculates the total reduced distance through collaboration after being informed that all delivery orders have been successfully fulfilled.
We follow this agent framework to develop the MACL system, a key component for the integrated platform.

%% Prototype
\section{An Integrated Collaborative Logistics Testbed}\label{sec:testbed}
While the MACL connects stakeholders in carrier collaboration together and ensure them work coherently, it is not inherently offer visibility into its operations, which is critical to a collaborative logistics testbed.
The digital twin technology can bridge this gap.
We thus proceed to discuss the development of an integrated environment {by combining the proposed MACL framework and digital twin technologies}. 
The resulting testbed implements a basic collaborative logistics digital twin, comprising a physical system with a scene map for deploying relevant physical assets and a digital counterpart that enables real-time monitoring and visualisation.
Unlike a purely digital system, this testbed includes both physical and digital elements. 
We will use ``the testbed'' in the rest of the paper to refer to this integrated collaborative logistics platform.
This section first describe its conceptual design and then details its concrete implementation and showcase.

%% Platform design
\subsection{Conceptual Design}\label{sec:design}
As discussed in \sectionautorefname~\ref{sec:introduction}, concerns of business privacy leakage and lacking trusted platforms and information hinder participation in collaborative logistics.
Therefore, the design of these platforms needs to ensure {\it neutrality}, {\it privacy} and {\it transparency}.
The platform should act in an unbiased, fair manner to maintain neutrality among all participating carriers.
Given that carrier collaboration involves a form of {\it coopetition}, where competing businesses cooperate for mutual interest, it is important to protect the independence and privacy of each carrier.
Moreover, the platform should ensure transparency for all stakeholders, thereby fostering trust among them.
Participants should have visibility into {their} operations and behaviour, with access to their state.
These principles are further concretised through the following specific design guidelines:
\begin{enumerate}
\item [1)] {\it Decentralised agent deployment}: 
    Carriers maintain independence by running their agents on their own machines.
    This decentralised approach allows carriers to control over their operations while engaging in collaboration.
    
\item [2)] {\it Collective resource management}: 
    Shared resources {such as facilities for running collaboration-related computations} are collectively managed instead of being privatised. 
    This ensures a fair and efficient distribution of resources among participants, enhancing collaboration and optimising resource utilisation.
    
\item [3)] {\it Real-time visibility}: 
    Stakeholders have access to real-time information into the platform's status.
    This feature empowers stakeholders to monitor and assess the platform's performance, enabling timely informed decision-making and proactive interventions.
\end{enumerate}
The {testbed} is designed in alignment with these {design} guidelines.

%% Figure: agent orgnisation
\begin{figure}
    \centerline{\includegraphics[width=0.60\textwidth]{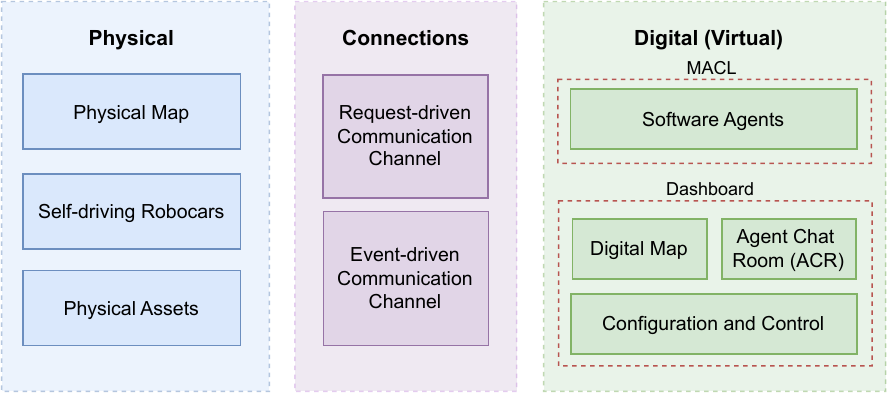}}
    \caption{{Main conceptual components of the testbed}.}
    \label{fig:main_components}
\end{figure}
%% Figure: context of diagram
\begin{figure}
    \centerline{\includegraphics[width=0.60\textwidth]{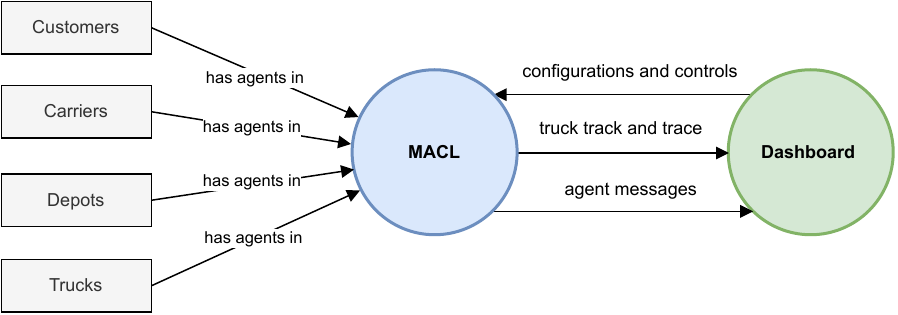}}
    \caption{{Context diagram of the testbed}.}
    \label{fig:system_context}
\end{figure}
\begin{figure}
    \centerline{\includegraphics[width=0.35\textwidth]{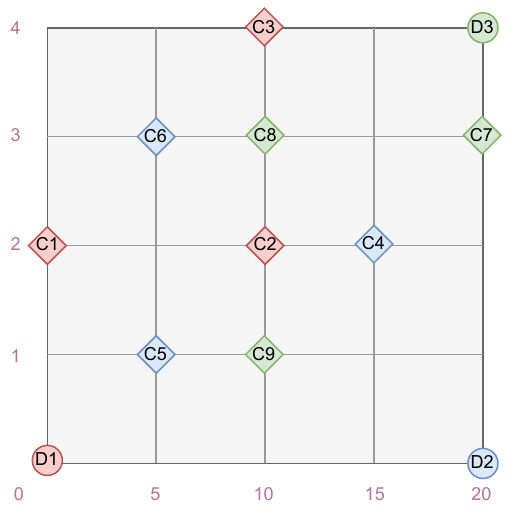}}
    \caption{
        Illustration of a 5\texttimes5 scene map. 
        The map has 25 available locations, including depots labelled as {\tt D1}-{\tt D3} and customers labelled as {\tt C1}-{\tt C9}.}
    \label{fig:scene_map}
\end{figure}

Following the mainstream three-layer design of digital twins\cite{tao2018digital}, the resulting testbed is conceptually composed of three main parts: a physical system, a digital system, and the connections between them. 
The main components within each of these parts are illustrated in \figureautorefname~\ref{fig:main_components}.
The physical system consists of a physical scene map, a fleet of self-driving robocars, and a set of physical assests.
The scene map depicts the physical environment where collaborations operate, as illustrated in \figureautorefname~\ref{fig:scene_map}.
There are two main physical assets: robotic cars (robocars) which simulate actual trucks used in transportation scenarios, and visual markers that denote the intersections and used for navigation on this map.
These robocars can autonomously fulfil transportation tasks by following predefined routes on {the scene map}.
Additionally, the map includes representations of carriers, depots, and customers, as shown in \figureautorefname~\ref{fig:scene_map}, simulating real-world configurations.
All of these physical assets can be represented by software agents, as previously discussed, acting on their behalf to interact with other assets.

The digital system is a virtual representation of the physical system, providing visibility into and control of its physical counterpart.
It consists of two main subsystems: the MACL system and the dashboard.
{While the MACL contains a set of software agents that represent relevant stakeholders in collaborative logistics, the dashboard comprises key digital components: a digital map, 
an agent chat room, and a configuration and control panel}. 
Its central component is the digital map --- a digital replica of the physical scene map.
This digital map updates itself regularly to synchronise with its physical counterpart. 
However, it only provides visibility into observable changes on the scene map, lacking transparency regarding the communications between the physical assets via their respective agents.
Agent chat room is thus added into the digital system, monitoring and displaying interactions between agents.
Through messaging, agents interact to collaborate or resolve conflicts, achieving coherence and collaboration.
This agent chat room visualises the entire messaging process, displaying messages and providing services for querying messages.
Additionally, a configuration and control panel that contains buttons to control and configure the physical system is added into the digital system.

We present a context diagram, as shown in \figureautorefname~\ref{fig:system_context}, to illustrate the inputs and outputs of the testbed from/to external factors. 
Within the digital system, the dashboard mainly controls and configures the MACL, and the MACL updates its state (e.g., trucks' locations and trace, agent messages) to the dashboard for visibility and monitoring. 
The external factors such as relevant stakeholders interact with the digital system through their representative agents running in the MACL. 
The MACL and dashboard together form the agent-based digital replica of the physical system.

The third part is the connections, comprising communication mechanisms and protocols that enable data exchange between the physical and digital systems.
We consider two distinct communication modes: request-driven and event-driven communication.
The first communication mode is stateless and unidirectional. 
This mode is commonly employed to retrieve resources or submit data. 
Examples include loading a web interface, querying data, or saving system configurations. 
This type of communication can be implemented either synchronously or asynchronously. 
The second is {driven by events}, offering bi-directional, low-latency, high-frequency communication services. 
Therefore, it is particularly suitable for real-time scenarios, such as real-time transmission of robocars' location data from the physical system to the digital system.

%% Figure: scene map

%% Platform development
\subsection{Implementation Overview}\label{sec:implementation}
The previous section describes the testbed's conceptual design, {with a focus on its main components}. 
This section presents its implementation details, including system architecture and concrete implementation of each component.

%% Architecture
\subsubsection{System Architecture}\label{sec:architecture}
\begin{figure}[!t]
    \centerline{\includegraphics[width=0.7\textwidth]{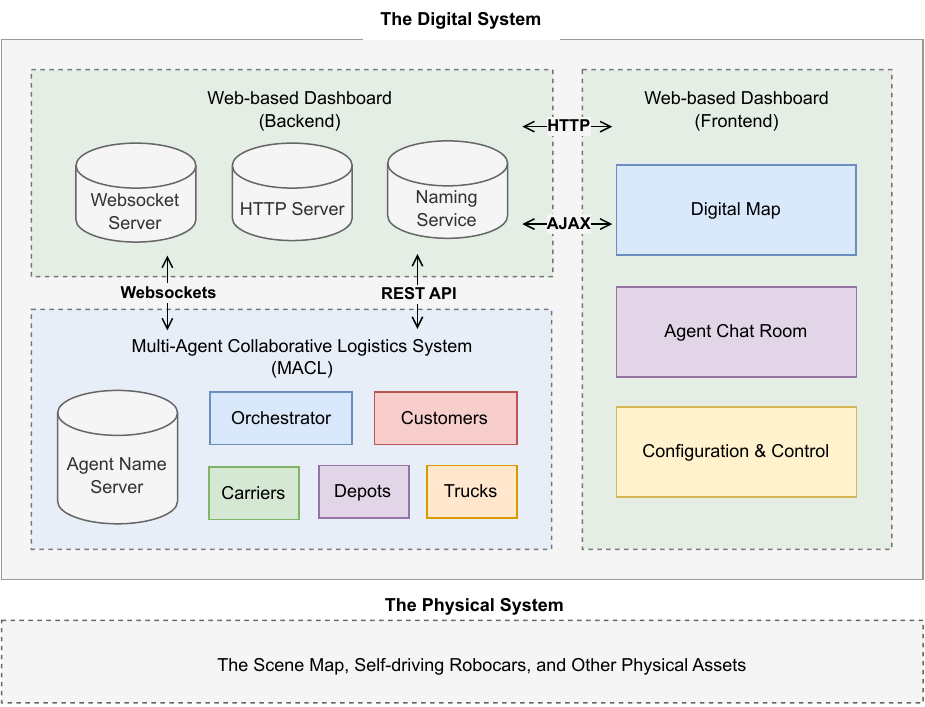}}
    \caption{System architecture of the testbed.}
    \label{fig:architecture}
\end{figure}

Architecturally, {besides the physical system that contains physical objects, the testbed consists of two main digital subsystems: the MACL system and the web-based dashboard}.
Adhering to the design guidelines presented in \sectionautorefname~\ref{sec:design}, these systems should be independent, loosely-coupled, and capable of running on geographically distributed machines. 
{We, therefore, design} a {{\it microservice} architecture for this testbed (see \figureautorefname~\ref{fig:architecture})}.

As shown in \figureautorefname~\ref{fig:architecture}, {the inter-service communication between the MACL and the dashboard are achieved via Websocket and RESTful API, allowing for both synchronous and asynchronous messaging between them}.  
The two systems are isolated from each other, and their communication occurs through messaging or API calls. 
This architectural choice ensures that the two systems can be independently deployed, operated, and maintained, without concerning about the other system's current state, internal logic, and implementation.
Specifically, the servers (e.g., agent name server, Websocket server, and HTTP server) and agents (orchestrator, carriers, etc.) can run on separate machines. 
Each of them has a distinct domain of responsibility, collaborating to establish a loosely-coupled system.

The physical system contains physical objects, such as the {physical scene map, self-driving robocars, and logistics stakeholders}.
Logistics stakeholders and robocars operate on the scene map, with corresponding agents acting on behalf of them in the digital system.
The digital system, apart from the MACL, includes a web-based dashboard designed for digitally twinning the physical system.
This dashboard can be logically separated into backend and frontend, as shown in \figureautorefname~\ref{fig:architecture}.
The frontend is connected to the backend through either synchronous (HTTP) or asynchronous (AJAX) requests, modifying its content or updating data to the backend. 
Both the frontend and backend contain additional components, which, in conjunction with others, will be further described in the {next} sections.

%% Figure: system architecture
%% ArUco marker examples
\begin{figure}[t]
    \centering
    \subfloat[ID: 0]{
        \includegraphics[width=0.12\linewidth]{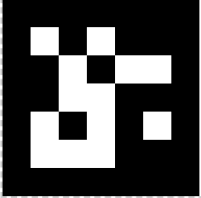}
    }
    \hspace*{-0.75em}
    \subfloat[ID: 1]{
         \includegraphics[width=0.12\linewidth]{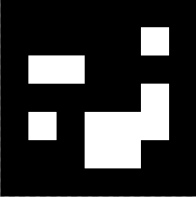}
    }
    \hspace*{-0.75em}
    \subfloat[ID: 2]{
        \includegraphics[width=0.1225\linewidth]{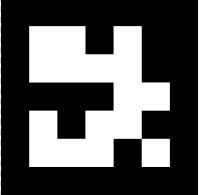}
    }
    \hspace*{-0.75em}
    \subfloat[ID: 3]{
        \includegraphics[width=0.12\linewidth]{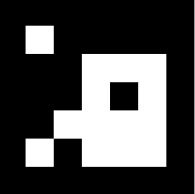}
    }
    \hspace*{-0.75em}
    \subfloat[ID: 4]{
        \includegraphics[width=0.12\linewidth]{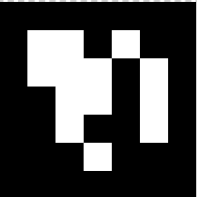}
    }
    \caption{Examples of ArUco markers (5\texttimes5 bits).}
    \label{fig:aruco_marker_examples}
\end{figure}
%% Figure: Testbed specification
\begin{figure}[t]
    \centerline{\includegraphics[width=0.55\textwidth]{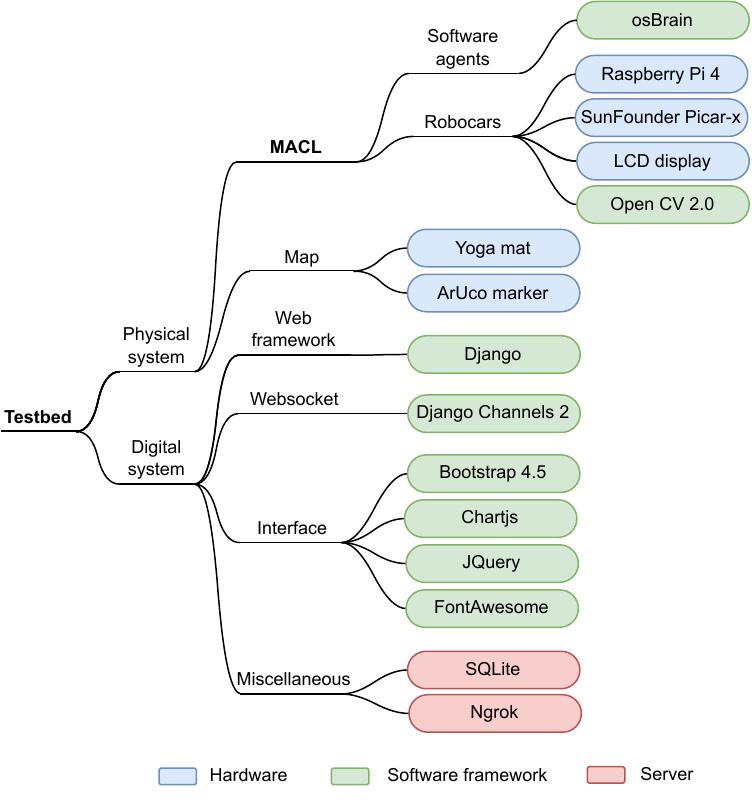}}
    \caption{Mindmap illustrating the testbed specification.}
    \label{fig:specification}
\end{figure}

%% Physical system
\subsubsection{Physical Map}\label{sec:map}
A physical environment, where relevant actors commonly operate, is crucial for the collaborative logistics platform. 
However, our goal is not to replicate the entire real-world environment, which is challenging, and yet unnecessary and uneconomical.  
Instead, we create a simplified yet effective simulated environment---a physical map with a coordinate system that allows localisation and navigation.
The map is divided into uniform grids, and its intersections are marked using ArUco markers \cite{garrido2014automatic}---a widely-used fiducial marker system in robotics (see \figureautorefname~\ref{fig:aruco_marker_examples} for the ArUco marker examples generated by the {\tt 5\texttimes5\_50} dictionary).
ArUco markers are binary square fiducial markers that contain unique identification numbers, which can be used to identify each intersection.  
Their detection is quick and simple, yet reliable and robust.
They can be generated automatically, have ample code space, and are free.
Therefore, ArUco markers are ideal for marking locations on this map, especially considering their compatibility with the low-cost onboard camera used for marker recognition on the robocars.

Moreover, given the map's division into a grid with designated IDs, this environment supports both localisation and navigation. 
Robocars can determine their locations on the map by identifying these markers and can find routes between two locations using basic {path-finding algorithms}. 
Grid density corresponds to quantity of accessible locations: denser grids generate more locations and achieve higher localisation accuracy.
Consequently, the grid combined with fiducial markers positioned at each intersection effectively converts the physical map space into a discrete two-dimensional coordinate system.

The map itself can be constructed from any materials with three main properties: non-slip, non-reflective, and portable.
After evaluating several options, yoga mats emerged as a suitable choice that satisfies the desired properties for the physical map.
They are also affordable and readily available for purchase.

%% Multi-agent collaborative logistics system
\subsubsection{MACL System}\label{sec:macl}
This MACL system mainly includes five types of agents, each exhibiting its own behaviour, as detailed in \sectionautorefname~\ref{sec:agents}. 
These agents can be categorised into two groups based on their mobility: mobile and stationary agents.
Among them, truck agents are mobile, driving trucks on the map to fulfil transportation requests, resulting in dynamic locations.
The rest of agents are stationary and has fixed locations on the map. 
To simulate the real-world logistics operations, we utilise self-driving robocars equipped with Raspberry Pi for running their onboard agents. 
These truck agents control the robocars' movements while facilitating direct communication with other agents to seek collaboration.
The self-driving of these robocars is achieved through a line-guided driving algorithm.
Stationary agents can be executed on various computing platforms, including desktop or portable computers, and single-board computers such as RaspBerry Pi.

Moreover, it is crucial that these agents can be run on different machines while interacting coherently; therefore, they must be capable of locating and communicating with one another via the network.
Therefore, the MACL requires to {enable agent} search and discovery (S\&D) service. 
This S\&D service provides agents with naming and connection functionalities, allowing them to locate other agents using memorable aliases rather than less meaningful URIs. 
Consequently, agents can easily establish connections through this service and communicate through messaging.
Agents receive messages and process them with corresponding handlers.
Effective handling messages requires recipients to understand both the structure and semantics of the received messages. 
This includes understanding sender information, message content, and potentially domain-specific data, such as delivery destination, lead time, and available delivery time window. 
Additionally, understanding semantics may need commonly-agreed protocols and shared ontologies among participating agents.

All the above aspects need to consider when developing a multi-agent system such as the MACL. 
To avoid starting development from scratch, we adopt the osBrain\footnote{\url{https://osbrain.readthedocs.io/en/stable/index.html}} framework to create agents in the MACL.
The osBrain, a Python-based, general-purpose MAS framework, was originally developed by OpenSistemas for a real-time automated trading platform.  
By leveraging osBrain, we can expedite the development process by distracting from low-level details and concentrating on the design and implementation of higher-level functionalities, such as defining agents' behaviour and their organisation and interaction.

%% Digital system
\subsubsection{The Dashboard}\label{sec:digital}
This section presents an overview of the implementation of the dashboard.
To provide convenient access to this system, we created a web-based digital system following the architecture presented in \sectionautorefname~\ref{sec:architecture}.
The developed system implements the three components (\figureautorefname~\ref{fig:main_components}): digital map, agent chat room, and configuration {and control} panel. 
We employed Websockets as communication channels for data exchange between the physical and digital systems.
This design ensures that events occurring in the physical system can be promptly reflected in the corresponding areas of the digital system.
Specifically, by establishing Websockets connections, the digital map and agent chat room allow for real-time monitoring of robocar movements and agent communications. 
The configuration component communicates with the physical system through request-driven communication channels, which can be implemented using techniques such as RESTful APIs and AJAX calls.

To facilitate development, Django\footnote{\url{https://www.djangoproject.com/}} was employed as the underlying web framework, in conjunction with other compatible plugins and frontend tools, to develop this digital system.
This system also incorporates auxiliary functionalities, such as data storage and naming service, for additional support.
The frameworks and tools used to develop this system is illustrated in \figureautorefname~\ref{fig:specification}.

%% Showcase
\subsection{Implementation Details}\label{sec:implementation_details}
%% Figure: The physical map and an example robocar
\begin{figure*}[t]
    \centering
    \subfloat[An assembled robocar.]{
        \label{fig:robocar}
        \includegraphics[width=0.45\linewidth]{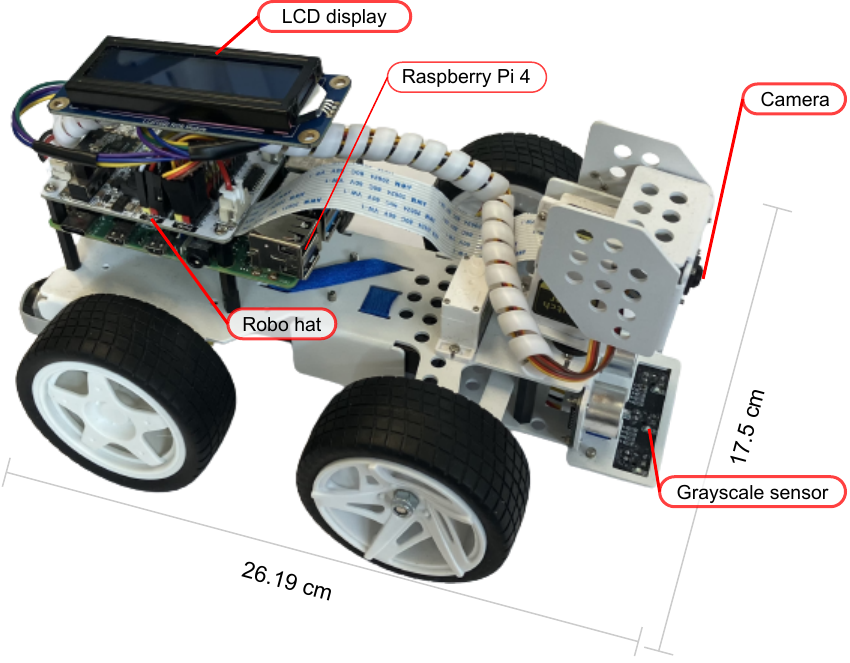}
    }
    \subfloat[The physical map]{
        \label{fig:physical_map}
        \includegraphics[width=0.375\linewidth]{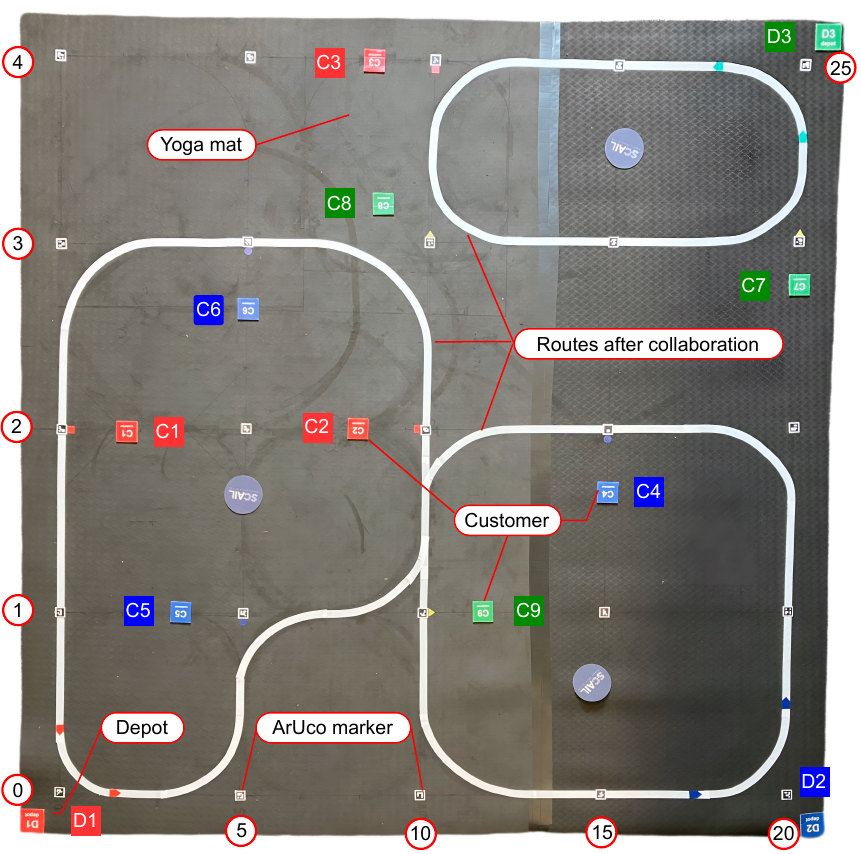}
    }
    \caption{Illustration of an assembled robocar and the physical map with annotations highlighting key components.}
    \label{fig:physical_assets}
\end{figure*}
Following the design described in the previous sections, we implemented a {\it demonstration} platform instead of a fully-fledged one, due to constraints in budget and physical space.
We details the implementation of the platform and presents a collaborative logistics showcase in this section.

%% Testbed setting up
\subsubsection{Environment Setting-up}
The main components that undergo simplification or reduction are associated with the physical elements, specifically the physical map and the self-driving trucks.
As described in \sectionautorefname~\ref{sec:map}, yoga mats meet the three desired {properties}.
We aligned two black yoga mats vertically and stitched them together to {create a {\tt 200\texttimes200}cm base material for constructing the map}.
For creating self-driving trucks, we utilised low-cost educational robocar kits, the SunFounder Picar-x, and Raspberry Pi as the toolkit for further development.

We assembled, configured, and tested multiple robocars for this showcase. 
Each robocar is equipped with a Raspberry Pi serving as its control centre and incorporates an LCD display to provide operational status updates.
Additionally, every robocar integrates a three-channel grayscale sensor and a low-cost camera, both used for detecting the environment ahead of the robocar.
An assembled robocar, with highlighted components, is shown in \figureautorefname~\ref{fig:robocar}.
Due to hardware limitations, these robocars cannot perform small, sharp turns. 
Considering {the limitations} of the map size and robocar maneuverability, we divided the map into a uniform {\tt 5\texttimes5} grid, resulting in a {\tt 5\texttimes5} coordinate system with 25 different locations (see \figureautorefname~\ref{fig:scene_map} for an illustration). 
Each location is identified by an ArUco marker with a unique ID ranging {from 0 to 24}. 
Each grid measures {\tt 45\texttimes45}cm, resulting in a {\tt 180\texttimes180}cm map area in the central region of the base map material.
This design ensures efficient utilisation of space while allowing for effective robocar movement.

\figureautorefname~\ref{fig:physical_map} shows the resulting physical map, which realises the conceptual map depicted in \figureautorefname~\ref{fig:scene_map}. As shown in \figureautorefname~\ref{fig:scene_map}, annotations in this figure highlight essential elements and provide explanatory details. 
And the numbers are the IDs assigned to the corresponding ArUco markers. 
The white lines on the map denote the designated driving routes, added for riving assistance.
This map provides a playground environment for investigating collaborative logistics.

%% Self-driving trucks
\subsubsection{Self-Driving Trucks}\label{sec:driving_algo}
Given the testbed's focus on collaboration, the trucks' self-driving capabilities are optional, not required. 
Any driving solution that successfully navigates from source to destination is acceptable.
We developed a simple yet effective self-driving algorithm, specifically a line-guided driving algorithm, to allow trucks to autonomously navigate designated routes and achieve localisation without manual intervention.
The algorithm uses the front grayscale sensor of the robocars to detect lines and adjust steering accordingly.
Additionally, the front camera of the robocars is utilised to identify the ArUco markers for localisation.
Based on whether the detected location ID corresponds to a target customer, the truck stops for delivery and its representative agent sends a notice-of-arrival message to the customer agent, or continues driving. 
We omit the technical details of this driving algorithm and hardware configurations due to the focus and space limitation of this article.

%% Development of the MACL
\subsubsection{Development of the MACL system}\label{sec:macl_development}
%% Figure: Interaction with the nds
\begin{figure}[t]
    \centerline{\includegraphics[width=0.5\textwidth]{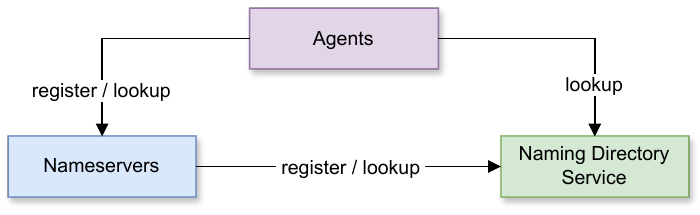}}
    \caption{Interaction flow enabled by the NDS.}
    \label{fig:nds_interaction}
\end{figure}
Following the MACL framework described in \sectionautorefname~\ref{sec:macl}, we implemented a MACL system using Python and the osBrain framework.
This implemented system consists of {\it sixteen} agents, including an orchestrator, three trucks, three depots (each with one truck), and nine customers.
While \figureautorefname~\ref{fig:agent_organisation} shows both carrier and depot agents for better representation of the problem, this implementation combines them together by integrating carrier agents' behaviours into their depot agents.
Additionally, we simplified the messaging process, reducing the number of interactions and minimising the data exchanged between agents. 
Messaging between agents is facilitated by the osBrain framework, specifically leveraging the ZeroMQ\footnote{\url{https://zeromq.org/}} messaging service to enable a request-reply communication pattern. 
This pattern enables effective agent message exchange and collaboration in this system.

We developed an agent nameserver using osBrain to provide search and discovery service for agents, enabling agents to conveniently find each other using assigned nicknames.
To ensure a consistent ``nickname'' for nameservers, we also implemented a naming directory service (NDS) for the digital system, {using} the REST architectural style.
The NDS maintains an lookup table containing a collection of nickname-IP address pairs, working as a ``phonebook'' for the MACL system. 
It allows an agents to submit a lookup request and retrieve its current IP address associated with the nameserver via its nickname, removing reliance on potentially dynamic IP address that may change at every system startup.
The NDS, in conjunction with the nameserver, allows the MACL system to be deployed on different machines under different networks.
Competing stakeholders can thus run and manage their own agents independently, which is in line with the second design principle discussed in \sectionautorefname~\ref{sec:design}.
The interaction flow enabled by this NDS is depicted in \figureautorefname~\ref{fig:nds_interaction}.

In this MACL system, three truck agents were deployed on three robocars.
These agents act as their decision-making bodies, control the robocars' driving and interacting with other agents to enable collaboration.

%% System running
\subsection{System Running and Showcase}\label{sec:system_running}
%% Figure: Dashboard initial screenshots
\begin{figure}[t]
    \centering
    \subfloat[Initial dashboard.]{
        \label{fig:dashboard_initial}
        \includegraphics[width=0.475\linewidth]{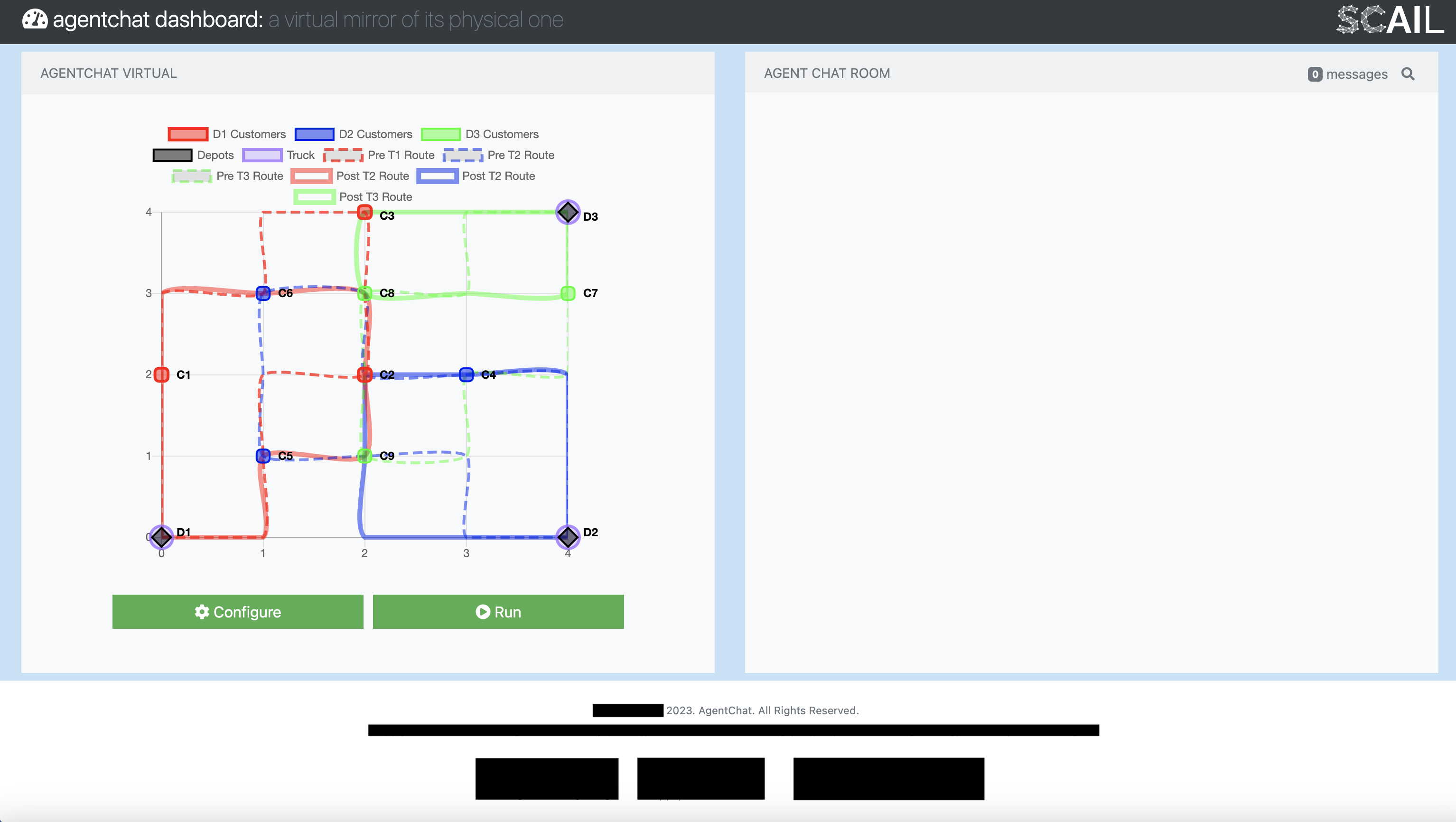}
    }
    \subfloat[Configuration popup.]{
        \label{fig:configure_popup}
        \includegraphics[width=0.48\linewidth]{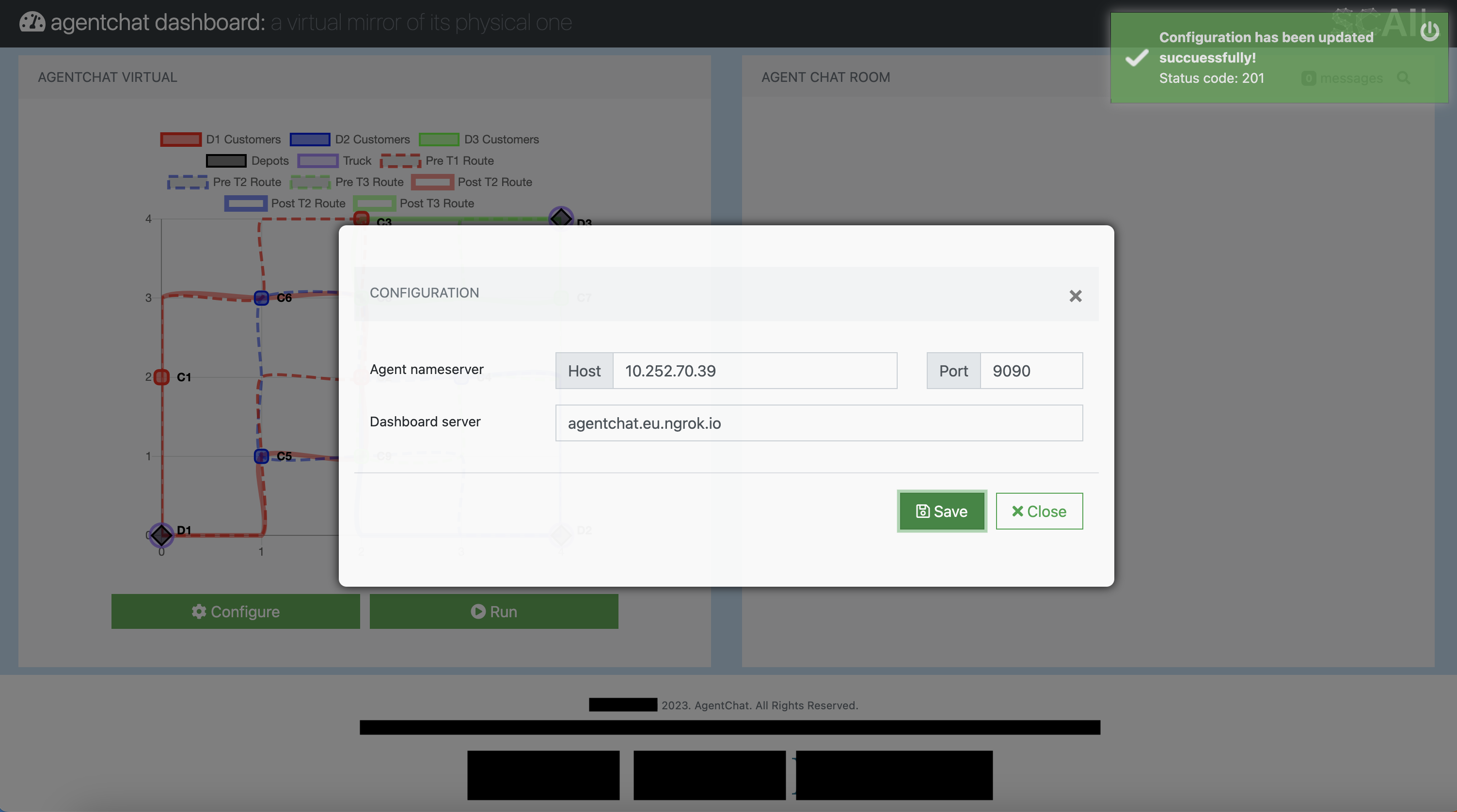}
    }
    \caption{Screenshots of the dashboard of the digital system at its initial and configuration status.}
    \label{fig:dashboard}
\end{figure}
%% Figure: Presteps to run the system
\begin{figure}
    \centerline{\includegraphics[width=0.55\textwidth]{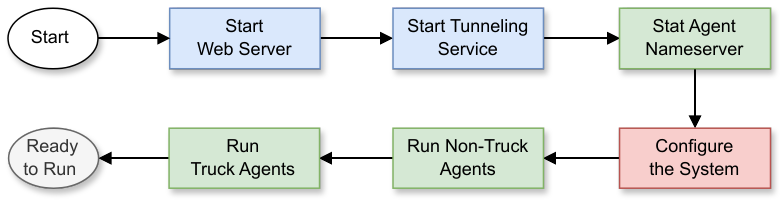}}
    \caption{Prerequisite steps to run the system.}
    \label{fig:run_system_steps}
\end{figure}

% Figure: status of the physical system and its mirror dashboard
\begin{figure*}[!t]
    \centering
    \subfloat[Start]{
        \label{fig:map_start}
        \includegraphics[width=0.40\linewidth]{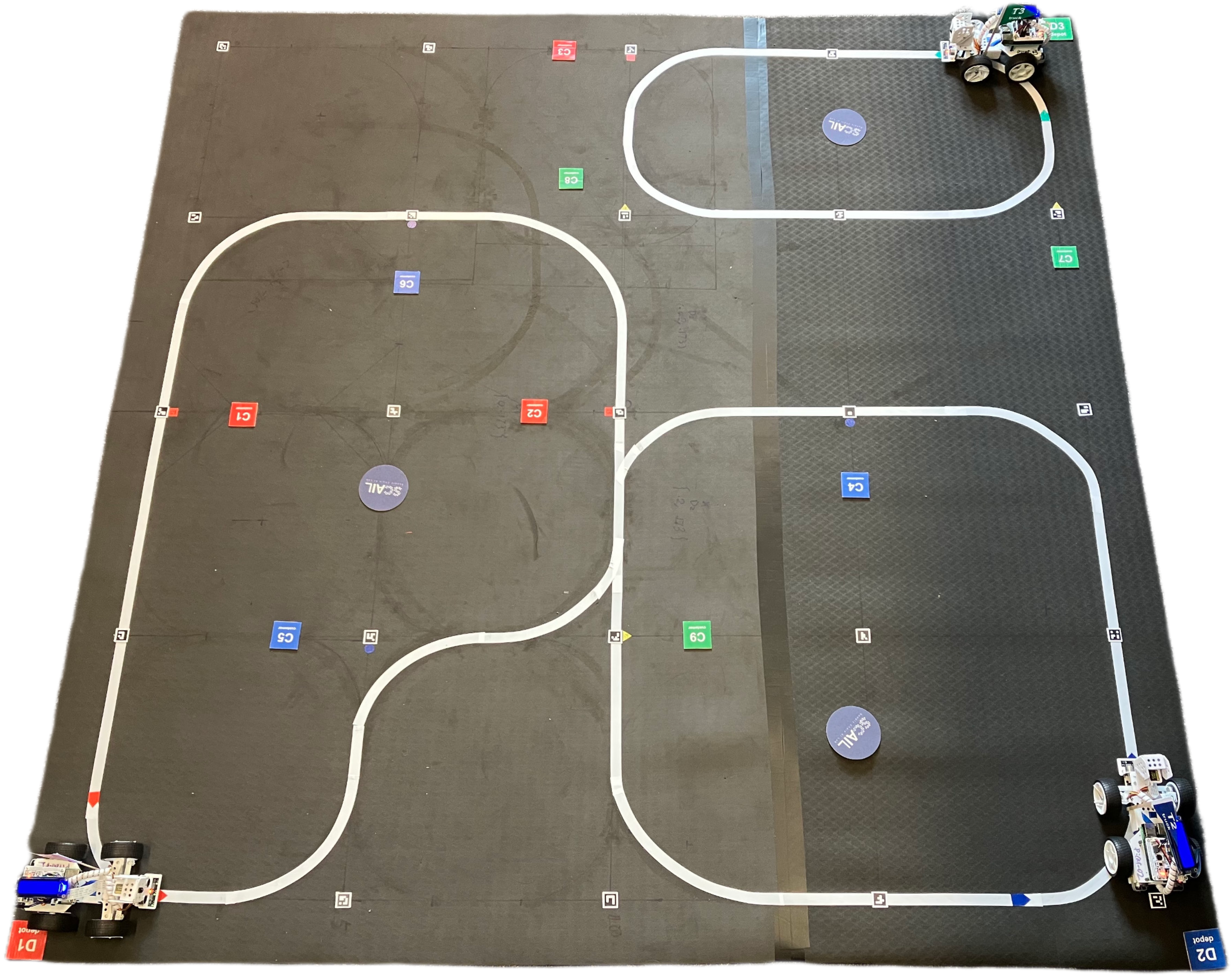}
    }
    \subfloat[Dashboard in the start of a run.]{
        \label{fig:dashboard_start}
        \includegraphics[width=0.55\linewidth]{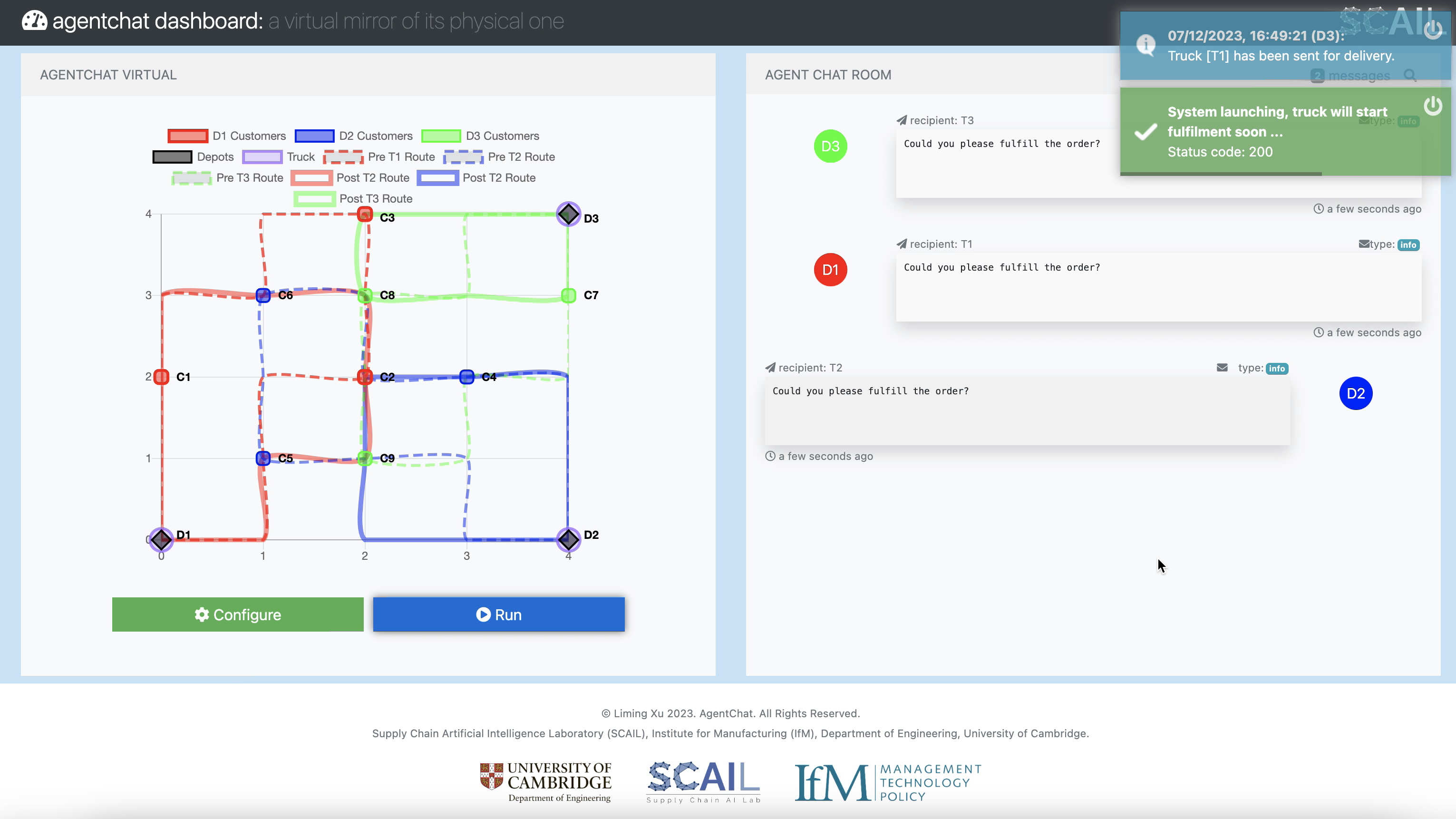}
    }
    \vfill
    \subfloat[Middle]{
        \label{fig:map_middle}
        \includegraphics[width=0.40\linewidth]{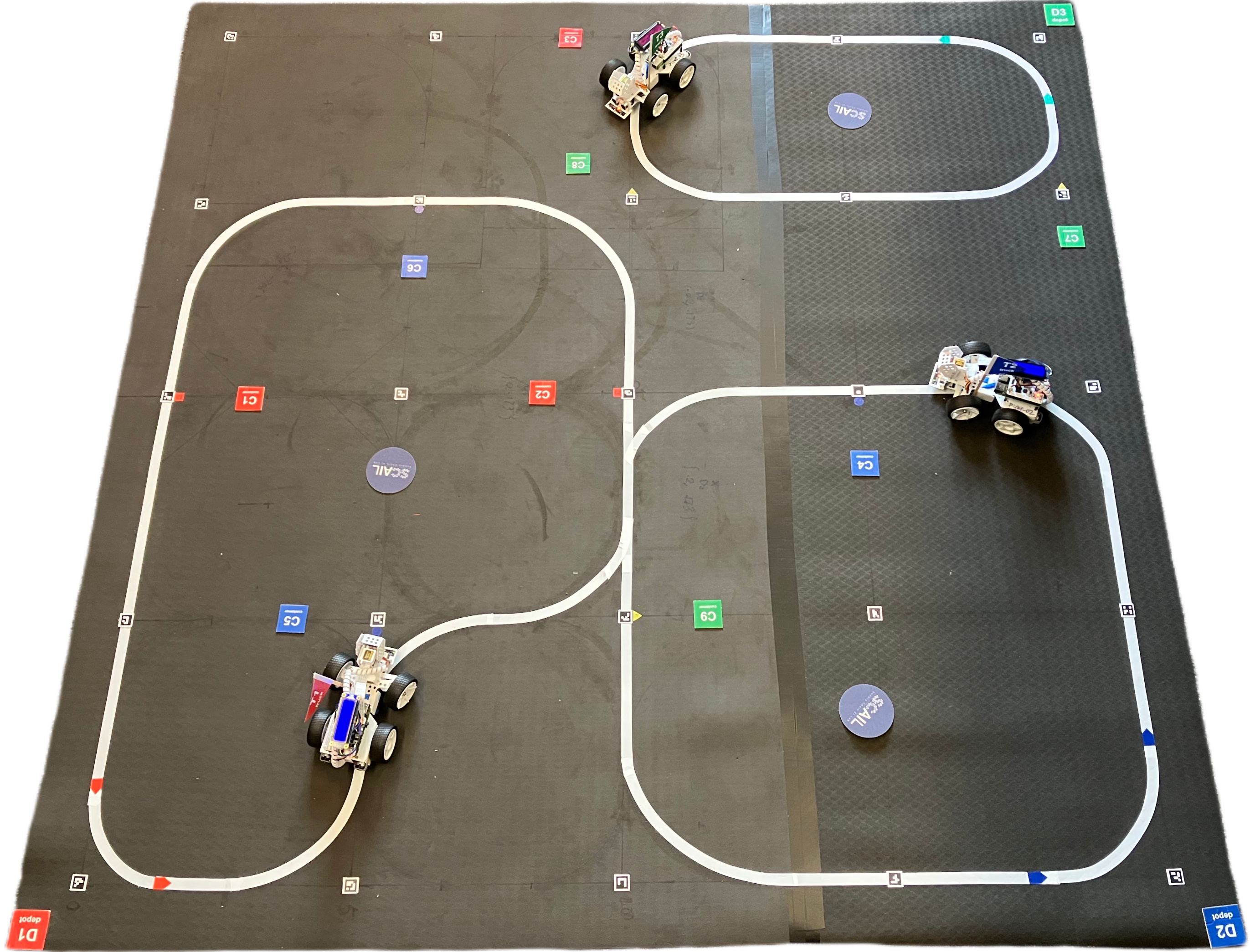}
    }
    \subfloat[Dashboard in the middle of a run.]{
        \label{fig:dashboard_middle}
        \includegraphics[width=0.55\linewidth]{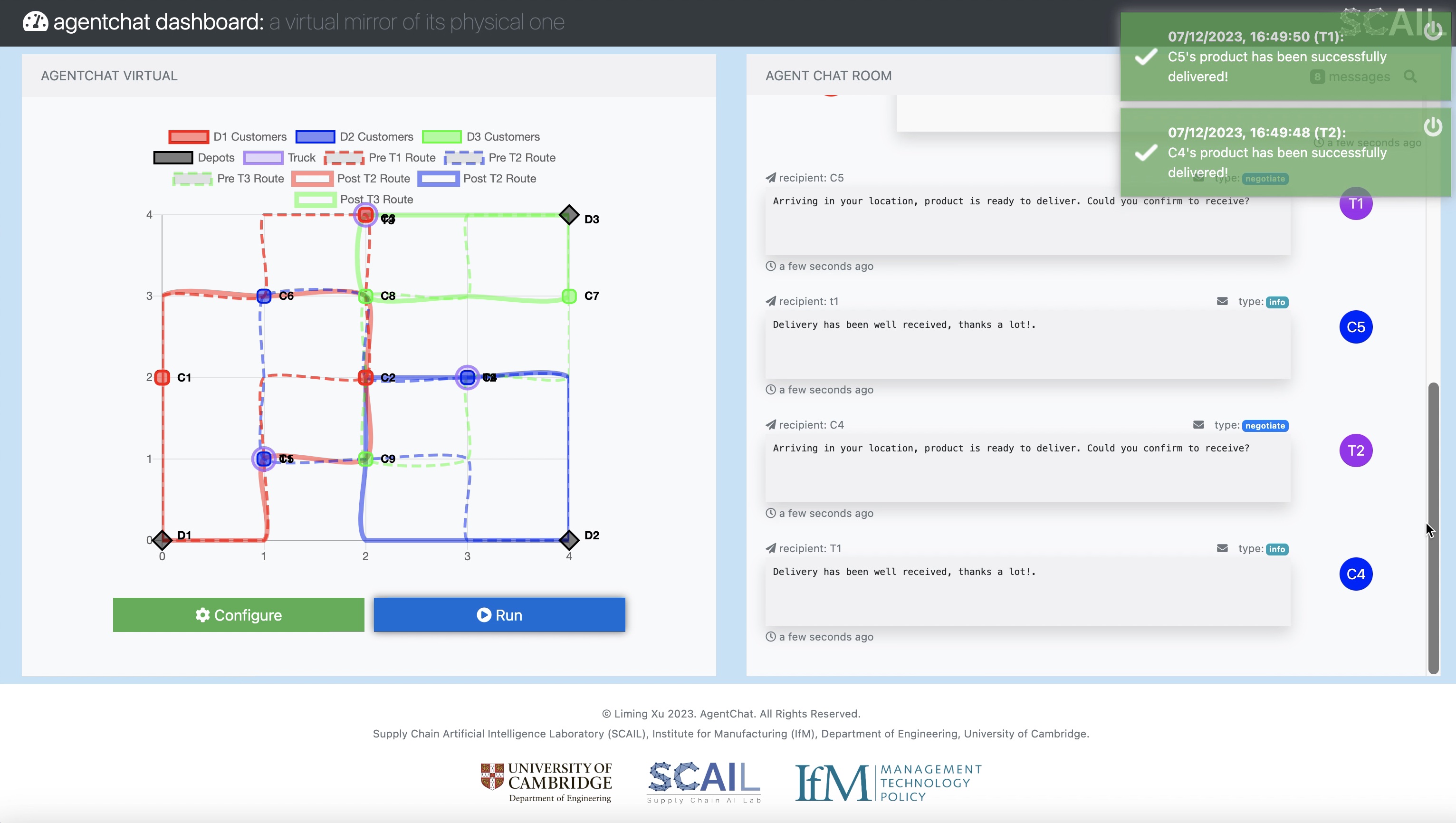}
    }
    \vfill
    \subfloat[End]{
        \label{fig:map_end}
        \includegraphics[width=0.40\linewidth]{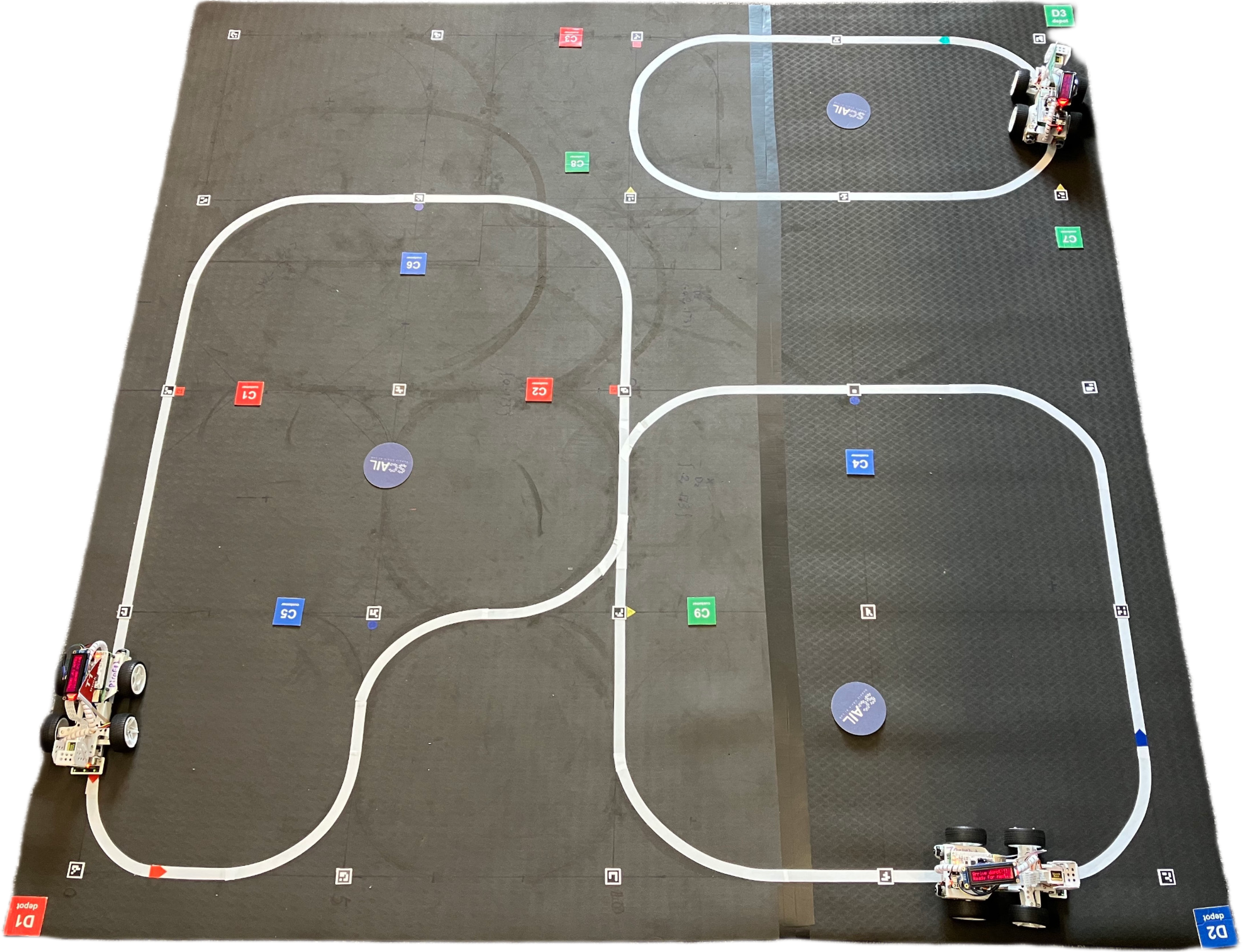}
    }
    \subfloat[Dashboard in the end of a run.]{
        \label{fig:dashboard_end}
        \includegraphics[width=0.55\linewidth]{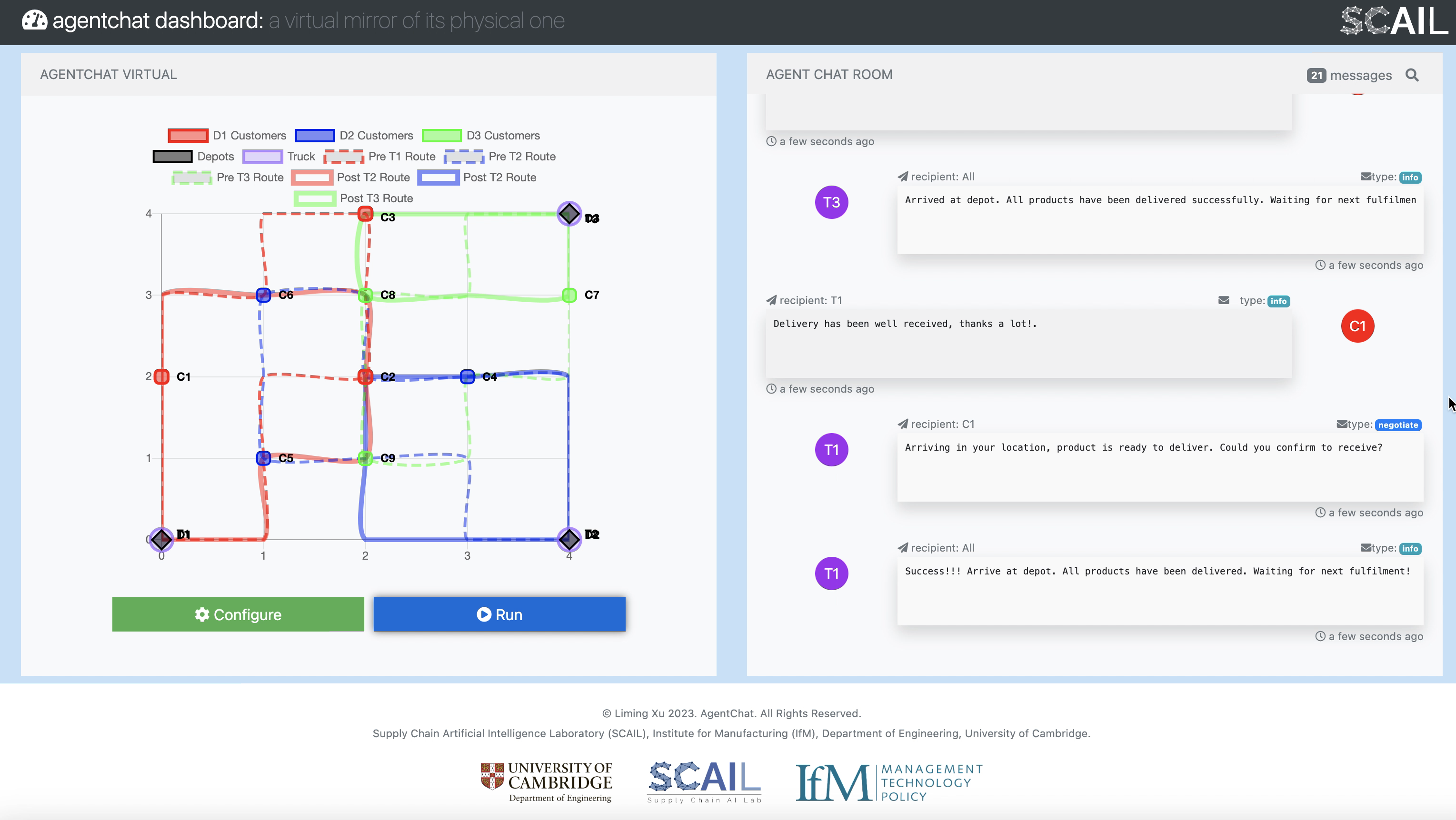}
    }
    \caption{
        Three states of the physical system (left) and theri corresponding screenshots of the dashboard of the digital system (right) at the start, middle, and end of a run.
    }
    \label{fig:system_running}
\end{figure*}
We implemented a web-based dashboard (see \figureautorefname~\ref{fig:dashboard}) for the digital system based on the design described in \sectionautorefname~\ref{sec:digital}. 
The dashboard consists of two panels: ``agentchat virtual'' and ``agent chat room''.
The left panel, agentchat virtual, displays a digital {\it replica} of the physical scene map (as shown in \figureautorefname~\ref{fig:physical_map}) on a 2-dimensional map. 
It mainly includes five digital elements: 
1) A 5 \texttimes~5 coordinate system, 
2) Labels representing entities relevant to collaboration, 
3) Routes of each truck before (dashed lines) and after (solid lines) collaboration,
4) Clickable legends for interactive showing or hiding of selected elements, and 
5) Control buttons for system configuration and launch. 
The right panel, agent chat room, initially appears empty with no messages until the system is running.

To run the system, a series of setup steps {\it must} be first performed, as illustrated in \figureautorefname~\ref{fig:run_system_steps}. 
These steps includes:
\begin{enumerate}
    \item Run the web server and allow {\it local} access to the interface via a browser. 
    
    \item Launch the tunnelling service powered by ngrok\footnote{\url{https://ngrok.com/}} to enable {\it remote} access to the digital system and its interface through a fixed URL\footnote{For example, in our tests, we are using\url{agentchat.eu.ngrok.io/dashboard/} as the URL to access dashboard.}.
    
    \item Start the nameserver after the web server and tunnelling service are operating.
    
    \item Update the nameserver's IP address (automatically detected) to the NDS via the configuration popup window as shown in \figureautorefname~\ref{fig:configure_popup}. 
    
    \item Finally, run all agents. Truck agents need to establish connections with corresponding {\it active} non-truck agents (e.g., customer agents) thus must be executed first.
\end{enumerate}

With all above steps completed, the system is ready for execution. 
When the ``Run'' button is clicked, the system starts to run.
The status of both the physical system and its digital counterpart at the start, middle and end of the running process is shown in \figureautorefname~\ref{fig:system_running} after post-collaboration routes are determined.
As shown in \figuresautorefname~\ref{fig:map_start}, at the beginning, the trucks (labelled {\tt T1}, {\tt T2}, and {\tt T3}) are parked at their {starting locations---the depots}: {\tt D1}, {\tt D2}, and {\tt D3}. 
These depots are located at the bottom left, bottom right, and top right corners of both the physical and the digital maps. 
The maps are populated with nine customers, evenly distributed over the area. 
Depots and customers are distinguished by coloured labels (red, blue, and green) that denote their association with specific trucks.
On the physical map, three white lines are marked to represent the routes after carrier collaboration.
These lines guide the self-driving of the trucks.
Ideally, these lines should be displayed only post-collaboration routes are determined rather than being shown at the beginning of a run.
The routes before carrier collaboration are not drawn on the physical map for clarity. 
However, on the digital map, both the routes before and after carrier collaboration {are displayed, denoted} by dashed lines and solid lines, respectively, as presented in \figureautorefname~\ref{fig:dashboard_start}. 
These routes with vectors of coordinates on the map are also showed in \tableautorefname~\ref{tbl:route_collaboration}.
When they receive messages containing delivery requests, the trucks illuminate their LCD displays in blue and adjust their front cameras to face down around 30 degrees, indicating their readiness to fulfil the requests. 
Messages and related notifications are displayed on both the agent chat room area and the upper right corner of the dashboard (see the right part of \figureautorefname~\ref{fig:dashboard_start}).

%% Figure: dynamic navigation redesign
\begin{figure}
    \centerline{\includegraphics[width=0.75\textwidth]{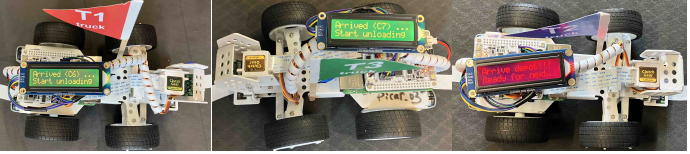}}
    \caption{Example messages displayed on the LCDs.}
    \label{fig:LCD_messages}
\end{figure}
%% Table: routes before and after collaboration
\begin{table}
\caption{{Routes before and after collaboration for the showcase.}}
\label{tbl:route_collaboration}
\centering
    {
    \begin{tabular}{c p{5cm} p{5cm} c} 
    \hline
    Truck No. & Route {\it before} collaboration & Route {\it after} collaboration & Distance Reduction \\ [0.5ex] 
    \hline\hline
    {\tt T1} 
    & 
    0$\rightarrow$5$\rightarrow$6$\rightarrow$11$\rightarrow$12$\rightarrow$13$\rightarrow$8$\rightarrow$3$\rightarrow$2$\rightarrow$1$\rightarrow$0
    & 
    0$\rightarrow$5$\rightarrow$6$\rightarrow$11$\rightarrow$12$\rightarrow$13$\rightarrow$8$\rightarrow$3$\rightarrow$2$\rightarrow$1$\rightarrow$0
    & from 12 to 10 blocks \\
    {\tt T2}
    & 
    20$\rightarrow$15$\rightarrow$16$\rightarrow$11$\rightarrow$6$\rightarrow$7$\rightarrow$8$\rightarrow$13$\rightarrow$12$\rightarrow$
    17$\rightarrow$22$\rightarrow$21$\rightarrow$20 
    & 
    20$\rightarrow$21$\rightarrow$22$\rightarrow$17$\rightarrow$12$\rightarrow$11$\rightarrow$10$\rightarrow$15$\rightarrow$20
    & from 20 to 8 blocks \\
    {\tt T3}
    &
    24$\rightarrow$19$\rightarrow$18$\rightarrow$13$\rightarrow$12$\rightarrow$11$\rightarrow$ 16$\rightarrow$17$\rightarrow$22$\rightarrow$23$\rightarrow$24
    & 
    24$\rightarrow$19$\rightarrow$14$\rightarrow$13$\rightarrow$18$\rightarrow$23$\rightarrow$24
    & from 10 to 6 blocks \\
    \hline
    \end{tabular}}
\end{table}

During delivery process, trucks use their cameras to scan the area ahread, identifying ArUco markers to determine their locations.
Upon reaching a customer' locations, they pause and send a notice-of-arrival message to the customer, waiting for a response.
Once receiving a confirmation-of-receipt message from the customer, trucks start a simulated offloading process, displaying a message on its LCD display (see \figureautorefname~\ref{fig:LCD_messages}) and flickering the display.
As shown in \figuresautorefname~\ref{fig:map_middle}, the three trucks are currently in the middle of fulfilling their delivery tasks.
Specifically, {\tt T1} and {\tt T2} have arrived at the location of their first customers, {\tt C5} and {\tt C4}, respectively, and have successfully delivered products to them.
{\tt T3} is approaching its second customer, {\tt C8}, but has not yet completed the delivery.
The digital system continually monitors and interprets the state of the physical system.
As illustrated in \figuresautorefname~\ref{fig:dashboard_middle}, trucks' location are constantly tracked and represented as coloured circles on the digital map.
The ongoing conversations between trucks and their respective customers are displayed in the right panel of the dashboard.
Additionally, milestone system events, such as {\tt T1} successfully delivered products to {\tt C5}, are promptly displayed via notifications showed at the upper right corner of the dashboard.

After a short stop following successful deliveries, trucks resume their journey, continuing along their assigned routes until all delivery assignments are fulfilled. 
Finally, they return to their respective depots. 
\figuresautorefname~\ref{fig:map_end} and \ref{fig:dashboard_end} show the state of the physical and digital system where all trucks have arrived their depots.
They send messages to their respective depots to inform their arrival and await for next fulfilment.
These interaction processes observed in this showcase verify the interaction flow presented in \figureautorefname~\ref{fig:agent_interaction}, showing the effectiveness of the proposed MACL framework.

Coherence among distributed agents in this system is achieved through inter-agent messaging. 
Trucks, customers, and depots communicate to coordinate their activities. 
Trucks also engage in dialogues to manage conflicts that might arise from simultaneous use of critical resources. 
For instance, when a truck is driving into a ``single-track road'' that accommodates only one vehicle at a time (like the road between coordinates (2, 1) and (2, 3)), the truck proactively informs nearby trucks about its road usage. 
Other trucks {have to wait} until the road is {available again}, preventing potential conflicts and enhancing efficiency.
This approach was employed in this testbed to avoid conflicts when using single-track roads on the map.

The demonstration shows the effectiveness of the proposal agent framework and its facilitation to {\color{red}enable digital twinning} for the given collaborative logistics scenario. 
We discuss the benefits, i.e., reduced total travel distance, achieved through the collaboration in this concrete showcase. 
As presented in \tableautorefname~\ref{tbl:route_collaboration}, the travel distances of the three trucks ({\tt T1}, {\tt T2}, {\tt T3}) have been reduced through collaboration from a minimum 2 blocks (i.e., grid unit of the map) to a maximum 12 blocks.
The total travel distance is reduced from 34 blocks before collaboration to 24 blocks after collaboration, which is approximately 29.4\% reduction in total distance, which is line with the existing claim in literature that collaboration would achieve one third cost reduction\cite{cruijssen2007horizontal}.
These reductions is achieved through less reduced overlapping routes of the three trucks due to  collaboration, which can be clearly observed from the \figureautorefname~\ref{fig:system_running} and \tableautorefname~\ref{tbl:route_collaboration}.
Such demonstrations of collaboration instances, stakeholders would intuitively observe the benefits of collaboration and raise their awareness of participation in collaboration. 
Additionally, this digital twin-based testbed provides for visibility into the processes such as collaboration orchestration, delivery fulfilment, negotiation, and message passing. 
It therefore would motivate participation in collaboration through transparency and simulations of collaboration scenarios. 
Although, due to the limitations of the physical map, the current testbed cannot be directly used for testing other collaboration instances without manual configurations on the map, which we will explore the extensibility in our future. 
However, the instance showcased here can represent many common scenarios in collaborative logistics.
The testbed represents an attempt to create an integrated hybrid platform beyond purely digital simulation, incorporating both physical and digital elements, stationary and mobile objects for examing collaboration logistics problem.

%% Discussion
\section{Discussion and Implications}\label{sec:discussion}
The showcase of carrier collaboration\footnote{This demo was accepted to be exhibited in AI UK 2023 (\url{https://ai-uk.turing.ac.uk/})---the UK's national showcase of data science and AI.}, despite its simplicity and directness, effectively demonstrates the suitability of the testbed for simulating or modelling of collaboration logistics scenarios.
The demonstration also provides a concrete example of how the integration of agents with digital twins could enable information sharing and build trust among carriers through the creation of a visible and transparent collaboration platform.

However, both the showcase and the testbed come with certain limitations. 
The showcased collaboration tackles a relatively small-scale problem, containing only {\it three} carriers, each only owns a single truck. 
While this three-carrier setup is a common practice in mainstream collaborative logistics research, it may not adequately capture the complexity of real-world scenarios involving large numbers of participants.
Additionally, the designed agent organisation is suitable for demonstration purposes, it may not be scale to larger collaboration scenarios. 
Consequently, the current testbed, along with its demonstrations, lacks the capability to study scalability-related problems.

The demonstration only includes a single collaboration instance, where pre- and post-collaboration delivery routes are predetermined and displayed on the map, as shown in \figureautorefname~\ref{fig:system_running} and \tableautorefname~\ref{tbl:route_collaboration}.
Therefore, this testbed is limited to demonstrating this specific ``fixed'' collaboration instance and requires physical configurations to explore more different collaboration instances.

These two limitations stem from the inherent constraints of the current testbed, namely, the limited area of its physical map and the lack of dynamic navigation capabilities.
While the first constraint is constrained by real-world conditions that might be beyond the scope of research efforts, the second constraint can be addressed by redesigning the navigation system of the physical map.
An attainable approach is to implement dynamic navigation by partitioning the map into uniform ``city blocks'' with white tap and then assigning each intersection a distinct ID. 
When determining the curvature and dimensions of these blocks, it is crucial to carefully consider the robocars' turning radius.
Our future work will explore this map design to build a dynamic navigation system for this testbed, accommodating more and arbitrary collaboration instances.

Despite these limitations, the testbed contributes not only to collaborative logistics but also to broader domains such as digital twins and Cyber Physical Systems (CPSs).
Its communication facilities enable distributed, decentralised entities to effectively discover and connect to each other in a dynamic IP allocation environment, adaptable to various distributed, multi-entity communication applications. 
While the system architecture in \figureautorefname~\ref{fig:architecture} is tailored for this testbed, it is somewhat general and can be used to build other digital twin systems and CPSs.
Essentially, the testbed exemplifies a form of CPS, combining digital and physical components to create intelligent, interconnected entities that facilitate collaborative logistics.

Moreover, this work aligns with the recent initiative for building the Physical Internet---an open global logistics system founded on physical, digital, and operational interconnectivity, achieved through encapsulation, interfaces and protocols \cite{montreuil2011toward}.
While the Physical Internet aims at fundamentally replacing current logistics systems with a combination of digital transportation networks, this work marks an early attempt towards this collaborative, hybrid, digital-driven next-generation logisitcs systems.
We will explore more on the convergence of collaborative logistics and the Physical Internet in our future work.

This testbed goes beyond conceptual explorations and pure digital simulations in existing studies (e.g., \cite{croatti2020integration}\cite{clemen2021multi}) by realising digital twin concepts through the MAS approach in collaborative logistics.
The testbed includes a physical collaborative logistics environment and its virtual replica. 
The digital system monitors and has the capability to control the operations of the physical system. 
However, the testbed currently only implements the collaborative logistics digital twin on the architectural aspects such as connectivity and interaction, lacking other essential functionalities of digital twins, such as advanced analytics and predictive capabilities.

%% Conclusion and future work
\section{Conclusion and Future Work}\label{sec:conclusion}
Collaborative logistics has garnered significant attention, largely driven by the ongoing green transitions.
However, carriers hesitate to participate in collaboration due to information sharing barriers and lack of integrated systems \cite{karam2021horizontal}. 
In this paper, we address these challenges by creating an integrated collaborative logistics testbed.
This testbed implements a hybrid system for carrier collaboration, capable of showcasing {collaboration instances} by the integration of agents with digital twin concepts.

Specifically, we presented an agent framework, called MACL, a simple yet effective approach to integrate distributed entities involved in collaboration. 
This framework consists of a set of representative or algorithmic agents collaborating to {achieve reduced total travel distance and better environmental and social good}.
We proposed a system architecture and employed it to implement an integrated collaborative logistics testbed. 
The testbed consists of a physical environment that includes a physical scene map, robocars, and other physical objects and a digital system that virtually replicates and monitors the physical environment. 
A demonstration of carrier collaboration involving sixteen agents, including three truck agents and nine customer agents, was conducted to showcase the effectiveness of the testbed.

Our future work will first address the limitations highlighted in the previous section.
This involves integrating a dynamic navigation system into the physical map, expanding its size to accommodate more available locations, and designing a scalable agent organisation. 
This will allow more robocars to operate concurrently and enable to study more collaboration instances.
Additionally, we will explore {\it decentralised} collaborative logistics by using trucks equipped with autonomous agents, where these agents can negotiate with each other to collaborate without a central orchestrator. 
This would be achieved by leveraging deep multi-agent reinforcement learning techniques \cite{mak2023fair} or integrating conventional MAS with emerging revolutionary foundation models \cite{bommasani2021opportunities,xu2024multi}.
Lastly, we will conduct studies to investigate the convergence of collaborative logistics and the Physical Internet, the two isolated but promising domains that facilitate transportation efficiency.

%% Reference
\bibliographystyle{IEEEtran}  
\bibliography{references}
\end{document}